\documentclass[floats,floatfix,amssymb,prd,superscriptaddress,nofootinbib,preprintnumbers,notitlepage,twocolumn]{revtex4-1}

\usepackage{subcaption}
\usepackage{ragged2e}
\DeclareCaptionJustification{justified}{\justifying}
\makeatletter
\newcommand{\subsetsim}{\mathrel{\mathpalette\subset@sim\relax}}
\newcommand{\subset@sim}[2]{%
  \vtop{\offinterlineskip\m@th
    \ialign{\hfil##\cr
      $#1\subset$\cr\noalign{\kern0.5pt}\scalebox{0.9}{$#1\sim$}\cr
    }%
  }%
}
\makeatother

\newcommand{\lmax}{\ell_{\rm max}}

\usepackage{amssymb,amsmath,verbatim,mathtools,needspace,enumitem,etoolbox,graphicx,physics,microtype,afterpage,bm}
\usepackage[dvipsnames, usenames]{xcolor}
\definecolor{linkcolor}{rgb}{0.0,0.3,0.5}
\usepackage{booktabs}
\usepackage[unicode, colorlinks=true, linkcolor=linkcolor, citecolor=linkcolor, filecolor=linkcolor,urlcolor=linkcolor, pdfusetitle]{hyperref}
\usepackage[all]{hypcap}
\usepackage[T1]{fontenc}
\usepackage[utf8]{inputenc}
\usepackage{tabularx}
\usepackage{float}
\usepackage{cleveref}

\usepackage{multirow}
\usepackage{pifont}
\usepackage{lmodern}
\usepackage{ulem}

\allowdisplaybreaks
\usepackage{tikz}
\usepackage{framed}
\usepackage{hyperref}
\definecolor{bluscuro}{rgb}{0.15, 0.2, .85}
\hypersetup{colorlinks, citecolor=bluscuro, linkcolor=black, urlcolor=bluscuro}

\newcommand{\beq}{\begin{equation}}
\newcommand{\eeq}{\end{equation}}
\newcommand{\barr}{\begin{eqnarray}}
\newcommand{\earr}{\end{eqnarray}}

\newcommand{\hp}{\hat{p}}

\newcommand{\be}{\begin{equation}}
\newcommand{\ee}{\end{equation}}

\newcommand{\T}{T_{\rm obs}}

\newcommand{\btheta}{\bm{\theta}}

\newcommand{\cern}{
CERN, Theoretical Physics Department,
Esplanade des Particules 1, Geneva 1211, Switzerland}

\newcommand{\iem}{
Instituto de Estructura de la Materia (IEM), CSIC, Serrano 121, 28006 Madrid, Spain}

\begin{document}

\title{
Cosmic Variance in Anisotropy Searches at Pulsar Timing Arrays
}

\author{Valerie Domcke}
\email{valerie.domcke@cern.ch}
\affiliation{\cern} 

\author{Gabriele Franciolini}
\email{gabriele.franciolini@cern.ch}
\affiliation{\cern} 

\author{Mauro Pieroni}
\email{mauro.pieroni@csic.es}
\affiliation{\iem} 

\begin{abstract}
Recent pulsar timing array (PTA) analyses show evidence for a gravitational wave background (GWB) with angular correlations consistent with the Hellings-Downs curve. Anisotropies are a key discriminator of the origin of this GWB, as they are expected to be at 1--20\% for astrophysical sources, but suppressed for cosmological GWBs.
However, contrary to gravitational wave detectors at higher frequencies, PTAs only take a few independent measurements of a GWB and consequently are highly sensitive to cosmic variance, which induces apparent anisotropies in individual realizations of an isotropic GWB. We demonstrate explicitly that statistical inference nevertheless remains robust, i.e., measurements are consistent with the underlying assumption of isotropy. This confirms that searches for anisotropies will be able to robustly discriminate astrophysical from cosmological GWBs.
En route, we demonstrate that the maximum multipole constrained by a PTA dataset scales linearly with the number of pulsars $\ell_{\rm max} \sim N_p$.
\end{abstract}

\maketitle

\preprint{CERN-TH-2025-159} 

\hypersetup{
colorlinks = true,
linkcolor=blue,
citecolor= blue, 
urlcolor=green!40!black,
}

\textbf{\textit{ Introduction. }}
Pulsar Timing Arrays (PTAs) monitor the arrival times of pulses from millisecond pulsars, whose intrinsic stability allows for precise predictions of their time-of-arrival at Earth. Gravitational waves (GWs) passing between Earth and these pulsars induce characteristic deviations, known as timing residuals, which are predictably correlated across pulsars. For an isotropic, gravitational wave background (GWB), the ensemble-averaged correlations trace the Hellings-Downs (HD) curve~\cite{Hellings:1983fr}, providing a distinctive signature for GW detection.

Recent results from worldwide PTA collaborations have reported evidence for a stochastic signal consistent with nanohertz GWs, displaying angular correlations in line with the HD expectation~\cite{NANOGrav:2023gor,Antoniadis:2023ott,Reardon:2023gzh,Xu:2023wog}. The precise origin of this background remains uncertain. While a GWB from supermassive black hole binaries (SMBHBs) is the leading interpretation, alternative scenarios involving a cosmological GWB from the early Universe remain compatible with current data~\cite{Madge:2023cak,NANOGrav:2023hvm,Antoniadis:2023xlr,Figueroa:2023zhu,Ellis:2023oxs}.

A promising avenue to distinguish these possibilities lies in the search for anisotropies in the GWB. Most cosmological sources predict highly isotropic backgrounds, whereas SMBHBs are expected to generate anisotropies at the $1-20\%$ level at low multipoles~\cite{Mingarelli:2013dsa,Taylor:2013esa,Mingarelli:2017fbe,NANOGrav:2023tcn,Sah:2024oyg,Sah:2024etc}. Thus, the detection of such anisotropies would provide strong evidence for an astrophysical origin of the signal, motivating the need to assess the sensitivity of current and future PTA datasets to these features.

In this work, we clarify the role of cosmic variance in PTA anisotropy measurements. Given their intrinsic frequency and angular coverage compared to the correlation length and time scales of the GWBs probed, PTAs are only sensitive to a few effective realizations of the GWB, unlike GW observatories at higher frequencies. Even in the limit of very many isotropically distributed pulsars and a GWB arising as the superposition of very many independent sources (e.g., of cosmological origin), this implies that the output of PTA measurements depends on the realization of the GWB in the local Universe, referred to as cosmic variance.\footnote{
Our terminology of cosmic variance is in analogy to CMB measurements and is referred to as sample variance in~\cite{Nay:2023pwu,Pitrou:2024scp}. It refers not only to the variance of the HD curve, but to all PTA observables. Additional scatter in the measurements arises from noise, from the finite number of pulsars in a PTA (also referred to as pulsar variance~\cite{Allen:2022dzg}) and, for a GWB due to SMBHBs, from their distribution in the sky (also referred to as shot noise~\cite{Grimm:2024lfj,Cusin:2025xle}). These effects arise in addition to the irreducible cosmic variance discussed here and are not the subject of this work.
} In particular, this implies that
locally measured anisotropies may significantly differ from ensemble averages.
Additionally, cosmic variance induces mild variations in the HD correlation itself, potentially influencing the sensitivity to anisotropies~\cite{Agarwal:2024hlj}.

Using realistic data simulations, we explicitly show that although maximum likelihood estimators obtained from individual realizations of an isotropic GWB can exhibit deviations due to cosmic variance, the recovered uncertainties render the inference statistically consistent with the underlying isotropic nature of the background, even in the signal-dominated regime.
This in particular implies that cosmic variance cannot `fake' anisotropies in a statistically significant way, and that the discrimination of astrophysical and cosmological GWBs based on anisotropies is robust. These results can be reproduced with \href{https://github.com/Mauropieroni/fastPTA}{\texttt{fastPTA}}, a publicly available Python code to forecast the constraining power of PTA configurations.

\vspace{2mm}
\textbf{\textit{ GWB anisotropies in PTA data. }}
PTAs measure the timing residual $\delta t_I$ induced by GWs along the line-of-sight between a pulsar $I$ (located in direction $\hat p_I$ at distance $D_I$) and Earth (located here at position $\vec 0$). The time shift induced at time $t$ then reads
\begin{equation}
    \delta t _I = 
   \frac{ \hat{p}_I^a \hat{p}_I^b}{2}
    \int_0^{D_I} {\rm d} s
    \, 
h_{ab} \left( t(s),\vec{x}(s) \right) 
     \label{eq:tim_res_gen} \,, 
\end{equation}
with $s$ denoting the affine parameter parameterising the geodesic, $t(s) = t - (D_I -s)$, $\vec{x} = (D_I -s) \hat{p}_I$, and $a,b$ the spatial components of the vectors. In the Fourier domain, the spin-2 perturbations of the metric read
\begin{equation}
\begin{aligned}
h_{ab}(t, \vec x) &=
\int_{S^2} {\rm d}^2\hat \Omega \, 
\int_{-\infty}^{\infty}{\rm d}f \,
e^{i2 \pi f(t - \hat \Omega \cdot \vec x)} \\
&\times 
\left[\tilde h_+(f,\hat \Omega) e^+_{ab}(\hat \Omega)
+\tilde h_\times(f,\hat \Omega) e^\times_{ab}(\hat\Omega)\right]
\,,
\label{e:hab(t,x)_stoch}
\end{aligned}
\end{equation}
where we introduced the polarization tensors $e^{+,\times}_{ab}(\hat k)$ with  $P = \{+,\times\}$.
Reality of $h_{ab}$ imposes $\tilde h_P(f,\hat \Omega) = \tilde h_P^*(-f,\hat  \Omega)$,
with $\hat \Omega$ denoting the normalized gravitational wave vector.
For a stationary and unpolarized GWB, we can express the ensemble average as
\begin{equation}
\langle \tilde h_P(f,\hat\Omega) 
\tilde h_{P'}^*(f',\hat \Omega') \rangle
= \frac{{\cal P} (f,\hat \Omega)}{4} 
\delta_{PP'}
\delta^2(\hat \Omega,\hat \Omega')
\delta(f-f')\,, 
\label{eq:hab_corr}
\end{equation}
where we take ${\cal P} (f,\hat \Omega)$ to be a deterministic function of the model. 
In this work, we assume the power spectral density (PSD) to be factorizable as
\begin{equation}\label{eq:spec_factorization}
{\cal P} (f,\hat \Omega) \equiv S_h(f) P(\hat \Omega) \, ,
\end{equation}
with normalization condition $\int {\rm d}^2 \hat \Omega P(\hat \Omega) = 1 $.
In general, the factorization in Eq.~\eqref{eq:spec_factorization} may not hold, as for the case of GWB produced by SMBHBs, for which anisotropies become larger at large frequencies \cite{Becsy:2022pnr,Sato-Polito:2023spo,Lemke:2024cdu,Gersbach:2024hcc,Raidal:2024tui,Moreschi:2025qtm}. However, this assumption does not affect our argument, and we will come back to this point in the conclusions.

The GW abundance today reads
\begin{equation}\label{eq:omegaGWdefS}
    \Omega_{\rm GW} h^2 = \frac{h^2}{\rho_c}\frac{ {\rm d} \rho_{\rm GW}}{{\rm d} \log f}
    =
    \frac{2 \pi^2 f^3}{3 H_0^2/h^2} S_h \, ,
\end{equation}
where $\rho_c$ is the critical energy density of the Universe and $H_0 /h = 1/(9.78 \,{\rm Gyr})$ is the Hubble parameter today. 
Building on the framework developed in previous studies~\cite{NANOGrav:2023tcn,Allen:1996gp,Anholm:2008wy,Mingarelli:2013dsa,Taylor:2013esa,Taylor:2015udp,Taylor:2020zpk,Ali-Haimoud:2020ozu,Ali-Haimoud:2020iyz,Pol:2022sjn}, we expand the GWB power spectrum in spherical harmonics and combine it with the geometric configuration of PTAs to predict the induced timing residual correlations.
To account for an anisotropic GWB, we expand $P(\hat \Omega)$ as
\begin{equation}\label{eq:PkSHexpansion}
   P( \hat \Omega )
    = 
    \sum_{\ell =0}^{\ell_{\rm max}}
    \sum_{m =-\ell}^{\ell} c_{\ell m} Y_{\ell m} (\theta, \phi) \, ,
\end{equation}
where $Y_{\ell m}(\theta, \phi)$ are the real-valued spherical harmonics. We adopt the same notation used in \cite{Depta:2024ykq}.
The normalization condition we use imposes $c_{00} = 1/\sqrt{4 \pi}$.

Given $N_p$ pulsars, the maximum number of multiples 
that can be constrained scales as $\ell _{\rm max} \propto N_p$ (not its square root, as reported in \cite{Romano:2016dpx,NANOGrav:2023tcn}). The argument proposed in \cite{Romano:2016dpx} is based on the assumption that from each pulsar data stream one can extract at most one complex number. This idea resonates with CMB temperature maps, where each pixel brings information about temperature or intensity of that portion of the sky alone (neglecting polarization, for simplicity). However, in the PTA case, each additional pulsar adds $N_p-1$ pulsar pairs, or new detectors, which are sensitive to the full sky through a unique pattern (given by the response function) that is not fully degenerate with the remaining pulsar pairs. We elaborate on this point in App.~\ref{app:lmax}.

\vspace{2mm}
\textbf{\textit{ Frequency domain correlation function. }}
The information on the GWB (anisotropies) is encoded in the correlations in the time delays measured between pulsars $I$ and $J$ at frequencies $i,j$, and
can be expressed as 
\begin{equation}
D_{IJ,ij} = 
\widetilde{ \delta t}_{I,i}
\widetilde{ \delta t}_{J,j}^* \, ,
\end{equation}
where  we introduced the Fourier space time delays
\begin{equation}
\begin{aligned}
\widetilde{\delta t}_{I,i} 
\equiv 
\frac{1}{i 2\pi}
\sum_P \int_{S^2} {\rm d}^2{\hat \Omega}
\int_{-\infty}^\infty 
\frac{ {\rm d} f}{f} 
\tilde{h}_{P}(f,\hat \Omega) 
R_I^{P}(f,\hat \Omega) 
\\
\, \times\text{sinc}[\pi(f_i - f)T_\text{obs}]\,,
\label{eq:sdef}
\end{aligned}
\end{equation}
and the response function
\begin{align}
R_I^{P}(f,\hat \Omega)
&\equiv
\frac{\hp^a_I \hp_I^b e^P_{ab}({\hat \Omega})}{2 (1 + \hp_I \cdot {\hat \Omega} )} 
\left(1-e^{-i 2\pi f D_I (1+\hat \Omega\cdot\hat p_I)}\right) \, .
\label{eq:response}
\end{align}

The ensemble average of $ D_{IJ,ij}$ can be derived using the statistical properties of the GWB \eqref{e:hab(t,x)_stoch}, which gives
\begin{equation}
\begin{aligned}
C_{IJ,ij} 
&= 
\frac{1}{24\pi^2}
\int_{-\infty}^\infty 
\frac{ {\rm d} f}{ f^2} \,
\Gamma_{IJ}(f) \,
S_h (f)  \\
&\times \text{sinc}[\pi(f_i - f)T_\text{obs}]\,
\text{sinc}[\pi(f_j - f)T_\text{obs}]\,,
\label{eq:exp_C}
\end{aligned}
\end{equation}
in terms of the overlap reduction function (ORF)
\begin{equation}\label{eq:ORF}
\Gamma_{IJ}(f)
\equiv
\int_{S^2}
{\rm d}^2 \hat \Omega \, 
\kappa_{IJ}(f,\hat \Omega) \,
\gamma_{IJ}(\hat \Omega)   \, 
P(\hat \Omega) \, ,
\end{equation}
with the pairwise timing response function $\gamma_{IJ}(\hat \Omega)$ \cite{Ali-Haimoud:2020ozu}
\begin{equation}
\begin{aligned}
\gamma_{IJ}(\hat \Omega) 
\equiv &
\frac{3}{4} 
\frac{\left[ \hat p_I \cdot \hat p_J 
- (\hat p_I \cdot {\hat \Omega}) 
(\hat p_J \cdot {\hat \Omega})\right]^2}
{(1+ \hat p_I \cdot {\hat \Omega})(1 + \hat p_J \cdot \hat \Omega)} + 
\\
 & -\frac{3}{8} 
(1 - \hat p_I\cdot {\hat \Omega})(1 - \hat p_J\cdot \hat \Omega) \, ,
\end{aligned}
\label{eq:gamma}
\end{equation}
and 
\begin{equation}
\kappa_{IJ} \equiv 
\left( 1 - e^{-2 \pi i f D_I (1  + \hat p_I \cdot \hat \Omega)}\right)
\left( 1 - e^{2 \pi i f D_J (1  + \hat p_j \cdot \hat \Omega)}  \right) \, .
\label{eq:kappa}
\end{equation}
For PTAs, the minimum frequency is limited by the observation time $f \sim 1/\T \sim  \text{nHz} \sim 0.1/\text{pc}$, while the typical distance to the pulsars is $D_I \sim~\text{kpc}$. Therefore, $f D_I \gg1$, so that the rapidly oscillating pieces in Eq.~\eqref{eq:kappa} are negligible when integrated over the directions $\hat \Omega$ (see also \cite{Mingarelli:2013dsa}), and $\kappa_{IJ}$ reduces to 
\begin{align}
    \kappa_{IJ} (f, {\hat \Omega} )\simeq 1 + \delta_{IJ} \, .
    \label{eq:kIJGIJ}
\end{align}
This simplification also implicitly assumes that fitting to the timing models reduces power below $1/\T$, see, e.g., Refs.~\cite{Blandford:1984xwb,Hazboun:2019vhv,Babak:2024yhu,Pitrou:2024scp}, for related discussion.
As a further test for the validity of this approximation, we will not adopt it for data generation, but only in the likelihood evaluation.
As a consequence of this and the initial assumption \eqref{eq:spec_factorization}, the correlation matrix is factorizable
\begin{equation}\label{C:factor}
 C_{IJ,ij}  = 
  C^p_{IJ} 
    \otimes
C^f_{ij} \, .
\end{equation}

\begin{figure*}
    \centering
    \includegraphics[width=.49\textwidth]{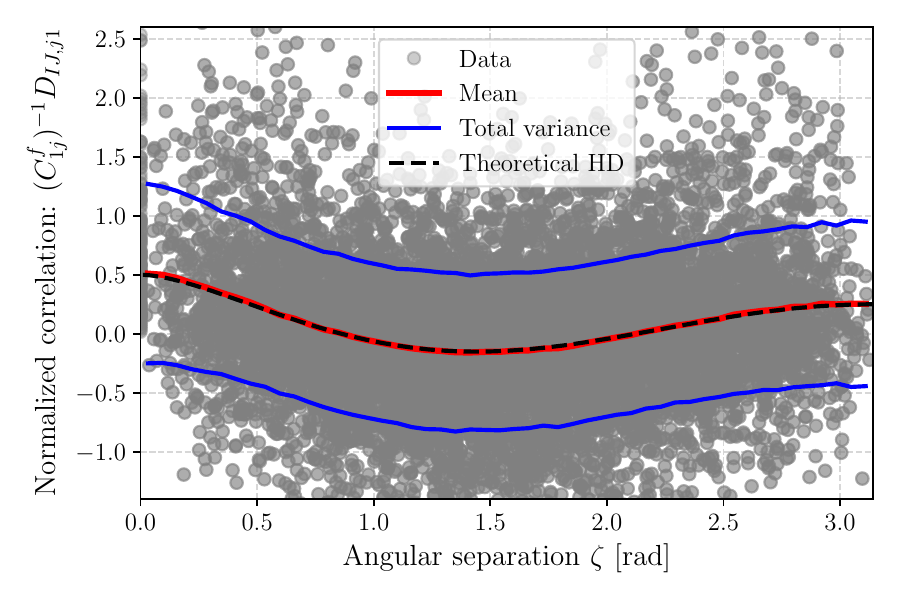}
    \includegraphics[width=.49\textwidth]{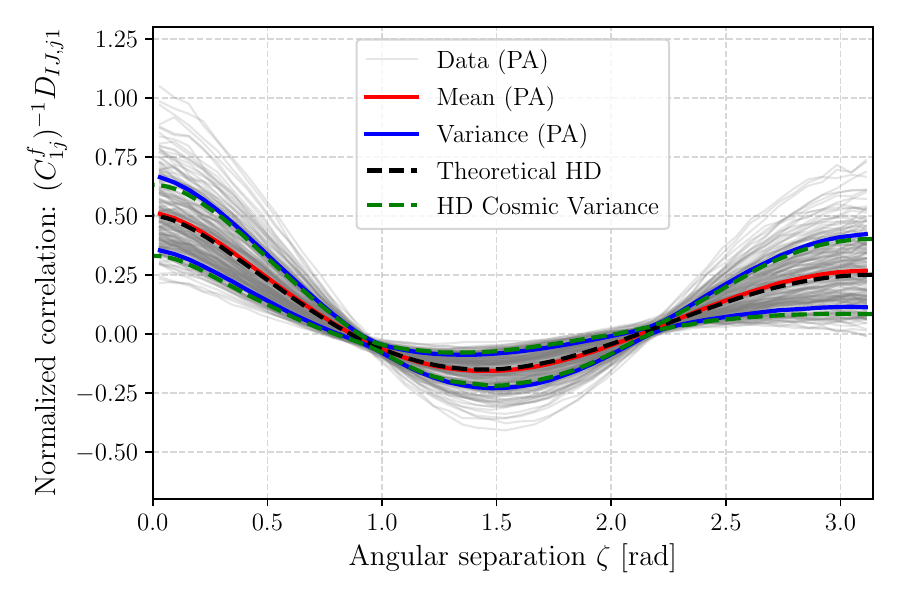}
    \caption{ 
{\it Left panel:} Correlations for hypothetical PTAs with $N_p = 100$ isotropically distributed pulsars, showing the first frequency bin $(i,j) = (1,1)$. 
Gray dots show $( C^f_{1j} )^{-1}D_{IJ,j1}$ for each of the $N_p (N_p + 1)/2$ autocorrelations (at zero angle) + cross-correlations in a single realization. Auto-correlations reach unity on average. The red line shows the mean over all pulsar pairs across 200 realizations, agreeing excellently with the HD prediction; the blue lines show the standard deviation.
{\it Right panel:} Grey solid lines show pulsar-averaged (PA) quantities (bins of size $\pi/50$) over 200 realizations of $N_p = 2000$ pulsars. Blue lines show the standard deviation after PA. The mean (red) agrees well with the HD prediction, and the variance agrees with the HD cosmic variance prediction (green dashed).
The black dashed line indicates the HD curve $\chi_{IJ}$ in both panels. We take $N_{\rm side} = 8$ and assume isotropic GWB. 
}
\label{fig:Cij}
\end{figure*}

Substituting \eqref{eq:PkSHexpansion} into \eqref{eq:ORF}, and using the orthogonality of $Y_{\ell m}(\theta, \phi)$, the ORF reads
\begin{align}
\label{eq:gamma_IJ_lm}
    \Gamma_{IJ} 
    =
    \sum_{\ell =0}^{\ell_{\rm max}}
\sum_{m =-\ell}^{\ell} c_{\ell m}  
\Gamma_{IJ, \ell m}\,.
\end{align}
The quantities $\Gamma_{IJ,\ell m}$, referred to as generalized ORFs~\cite{Mingarelli:2013dsa}, encode the geometric structure of the PTA experiment. They can be viewed as a generalization of the HD function $\chi_{IJ} \equiv c_{00}\Gamma_{IJ, 00}$~\cite{Hellings:1983fr}, see Fig.~\ref{fig:Cij} (which is only a function of the angle  $\cos \zeta_{IJ} = \hat{p}_I \hat{p}_J$ between pulsars $I$ and $J$ as dictated by symmetries, see, e.g.,~\cite{Kehagias:2024plp}), to higher multipoles.

\vspace{2mm}
\textbf{\textit{ Data simulation. }}
We start from Eq.~\eqref{eq:sdef} and turn the Fourier space integration into a sum over a discrete set of frequencies $N_f^{\rm int}$, as well as the angular integration over a pixelized sky with $N_{\rm pix}$ pixels to obtain
\begin{align}
\widetilde{\delta t}_{I,i} &= 
\sum_{P,n,k}
\left [
\tilde{h}_{P,n,k}
R_{I,n,k}^{P}
s_{ik}^-
-
\tilde{h}^*_{P,n,k} R_{I,n,k}^{P*}
s_{ik}^+
\right ]
\frac{\Delta\hat\Omega \Delta f }{2\pi f_k} 
\,,
\label{eq:Ctimesim}
\end{align}
where we introduced 
\begin{equation}
    s^{\pm}_{ik}
    \equiv 
    {\rm sinc}(\pi(f_i \pm f_k)\T) \, ,
\end{equation}
and the pixel index $n$ labels directions on the sky using an equal-area tessellation with angular size $\Delta \hat \Omega = 4 \pi / N_{\rm pix}$ obtained with the \texttt{HEALPix} package~\cite{Gorski:2004by}, while 
the index k labels positive only frequencies $f_k = f_{\rm min} + k \Delta f$, with $k=0, 1, ... , N_f^{\rm int}$. 
We take $f_{\rm min} = 1/ 2 \T$ and $\Delta f = 1/ 10 \, \T$\footnote{The low frequency integration cut-off is motivated by the effective sharp reduction of power at low frequencies induced by fitting the timing model ~\cite{Blandford:1984xwb,Hazboun:2019vhv,Babak:2024yhu,Pitrou:2024scp,Allen:2024bnk}.
We tested that our choice does not significantly affect our results.
Choosing a lower $f_{\rm min}$ and very red spectra would require going beyond the 
approximation \eqref{eq:kIJGIJ} in the likelihood, as corrections proportional to $f_{\rm min} D_I$ would become relevant for large $N_p$ networks.}. We perform the analysis for a base of positive frequencies $f_i = i/\T$, with $i>0$, therefore
\begin{equation}
D_{IJ,ij>0} = 
2{\rm Re} \left [
\widetilde{ \delta t}_{I,i}\widetilde{ \delta t}_{J,j}^*
\right ] \, .
\end{equation}

To generate GWB realizations, we simulate the complex coefficients $\tilde{h}_{P,n,k}$ appearing in Eq.~\eqref{eq:Ctimesim} as Gaussian random variables with zero mean and two-point function
\begin{equation}
\begin{aligned}
    \langle \tilde h^{*}_{P,n,k}  \tilde h_{P',n',k'} \rangle 
    &= \delta_{PP'}\delta_{nn'}\delta_{kk'}
    \frac{H_{n,k}}{\Delta\hat\Omega\,\Delta f} \, , \\
    \langle \tilde h_{P,n,k} \tilde h_{P',n',k'}\rangle 
    &=
    \langle \tilde h^{*}_{P,n,k} 
    \tilde h^{*}_{P',n',k'}\rangle 
    = 0 \, .
    \label{eq:covariance_pix}
\end{aligned}
\end{equation}
This equation is the discretized version of Eq.~\eqref{eq:hab_corr}, with 
$H_{n,k} = P(\hat \Omega_n)S_h(f_k) /4$,
and $H_k = \sum_n H_{n,k} = S_h(f_k)/4$
is directly related to the sky-integrated strain PSD in the \(k\)-th frequency bin\footnote{Notice that $\Delta \hat \Omega$ in Eq.~\eqref{eq:covariance_pix} ensures that the total power in a given sky patch is pixelization-independent.\label{footnote:pixelization}}. 
 When assuming anisotropic injections with $c_{\ell>0,m} \neq 0$, we distribute power across pixels according to \eqref{eq:PkSHexpansion}.
Because these variables are complex and Gaussian, we take the real and imaginary parts of \(\tilde h_{P,n,k}\) 
as
$\tilde h_{P,n,k} = (h^r_{P,n,k} + i h^i_{P,n,k})/\sqrt{2}$, 
where \(h_r\) and \(h_i\) are independent real Gaussian random variables with zero mean and variance
$\sigma_{n,k}^2 = H_{n,k} / {\Delta\hat\Omega\Delta f}$.

The expectation value of $D_{IJ,ij}$ \eqref{eq:exp_C} becomes
\begin{align}\label{eq:Cijens}
    C_{IJ,ij>0} = 
    \frac{4}{3} \Gamma_{IJ} 
   & 
   \sum_{k} \frac{\Delta f H_k}{4 \pi^2 f_k^2}
  \left [ s^-_{ik} s^-_{jk} + s^+_{ik} s^+_{jk}  \right ] \, ,
\end{align}
with
\begin{equation}
\Gamma_{IJ} = C^p_{IJ} = 
\frac{3}{2} 
\Delta \hat \Omega 
\sum_{P,n} H_n \, R_{I,n,k}^{P} \, R_{J,n,k}^{P*,} \, ,
\end{equation}
where, assuming the factorization \eqref{eq:spec_factorization}, i.e. $H_{n,k} = H_{n} H_{k}$, holds, \eqref{eq:kIJGIJ} implies the l.h.s. to be frequency independent.

In our forecasts, for each realization, we generate pulsar positions randomly on the sky, assuming a uniform distribution over the sphere. We illustrate these realizations in Fig.~\ref{fig:Cij}, showing the normalised $D_{IJ,ij}$ as a function of the pulsar-pulsar angular separation.
On the left, we display the total variance computed over all pulsar pairs and frequencies. On the right, we perform the pulsar average by binning according to angular distances and averaging within each bin. The mean closely follows the HD prediction, while the scatter agrees with the expected cosmic variance \cite{Allen:2022dzg}.
The latter is invariant under the increase of the number of pulsars, reflecting the irreducible variance arising from the GWB realization in our Universe.

\vspace{2mm}
\textbf{\textit{ Inference. }}
In this work, we focus on the strong-signal regime and neglect the role of instrumental and intrinsic noise. While technically straightforward, incorporating noise in the analysis would primarily reduce the effective signal-to-noise ratio without qualitatively altering our conclusions (see Ref.~\cite{Babak:2024yhu} for an in-depth discussion). In practice, 
we neglect the noise contribution, $C_{n,IJ,ij}$, to the total cross-correlation introduced in~\cite{Babak:2024yhu}, 
and the full ensemble average covariance matrix for the timing residuals from pulsars $I,J$ and frequencies $i,j$ takes the compact form \eqref{eq:Cijens}.
This covariance matrix extends the one considered in Ref.~\cite{Babak:2024yhu} as it accounts for frequency-frequency correlations and an arbitrary GWB angular dependence, i.e., it does not rely on the assumption of an isotropic GWB. We stress again that we rely on the factorization in \eqref{C:factor}. Under this assumption, since frequency and pulsar indices do not mix, we can neglect the transmission function from the analysis without affecting the conclusions regarding anisotropies.

Assuming the data $\tilde{\delta t}_{I,i}$ to be Gaussian variables with zero mean and described only by their variance $C_{IJ,ij}(\bm \theta) $, the log-likelihood can be written as
\cite{Moran1951HypothesisTI,Contaldi:2020rht,Bond:1998zw,Babak:2024yhu}
\begin{equation}\label{eq:likelihood}
- \ln \mathcal{L}
(\tilde{\delta t} \vert \btheta )
=
\sum_{IJ,ij>0} 
 \ln [\det C_{IJ,ij}] 
+
\tilde{\delta t}_{I,i} C^{-1}_{IJ,ij}\tilde{\delta t}^*_{J,j}  \, ,
\end{equation}
where we have ignored additive terms that are independent of model parameters $\btheta$, which includes only the coefficients $c_{\ell m}$ parametrizing GWB anisotropies.
We do not infer the GWB spectral properties, as, thanks to the factorizability \eqref{C:factor}, they are irrelevant for analyses of anisotropies in the strong-signal regime.
To speed up the inference, we adopt an iterative procedure (e.g., \cite{Franciolini:2025leq}).
We find the maximum likelihood (ML)
estimators for 
$\hat \btheta \equiv \hat c_{\ell m}$ in \eqref{eq:PkSHexpansion}  
by iteratively performing a gradient descent with a Fisher Information Matrix (FIM)-informed step size, i.e. 
$\hat{\btheta}_\alpha^{(i+1)} = 
\hat{\btheta}_\alpha^{(i)} + \delta\hat{\btheta}_\alpha^{(i)}$ with 
\begin{align}
& \delta{\hat \btheta}_\alpha^{(i)} = 
\sum_{\beta}[F^{-1}(\btheta)]_{\alpha \beta}
\frac{\partial \ln{\cal L}(\tilde \delta t \vert \btheta)}{\partial\btheta_\beta}
\Bigg|_{\btheta={\hat \btheta}^{(i)}} \, ,
\label{eq:iterative}
\end{align}
where the partial derivative of the log-likelihood reads
\begin{equation}
\frac{\partial \ln{\cal L}(\tilde \delta t \vert \btheta)}{\partial\btheta_\alpha}
= 
{\rm Tr}\left[C^{-1}\frac{\partial C}{\partial\btheta_\alpha}C^{-1}(D-C)\right],
\end{equation}
with matrix multiplication and trace over on both pulsar $IJ$ and frequency $ij$ indeces (omitted to avoid clutter). These steps are repeated until the changes \( |\delta\hat{\btheta}_\alpha^{(i)}| \) become smaller than a predefined convergence threshold, which we set to be $1\%$ of the current parameter uncertainty estimated from the FIM 
defined as (e.g.,~\cite{Bond:1998zw})
\begin{equation}
F_{\alpha \beta} 
= 
 {\rm Tr}\left[ 
 C^{-1}
\frac{\partial C}{\partial \btheta_\alpha}
C^{-1}
\frac{\partial C}{\partial \btheta_\beta} \right] \,.
\label{eq:fisher}
\end{equation}
Assuming the posterior distribution is nearly Gaussian, one can obtain the measurement uncertainties, as well as the model parameter correlations, by inverting the FIM (see, e.g.,~\cite{Coe:2009xf}). 
We validate the minimization procedure by also performing likelihood minimization with other numerical routines, confirming our findings. Also, we validated the FIM procedure by running an MCMC to confirm both ML and FIM estimates presented below. Finally, the accuracy of the inference is demonstrated by the P-P plots we report in App.~\ref{app:pp-plot}.

\begin{figure*}[t]
    \centering
     \includegraphics[width=1\textwidth]{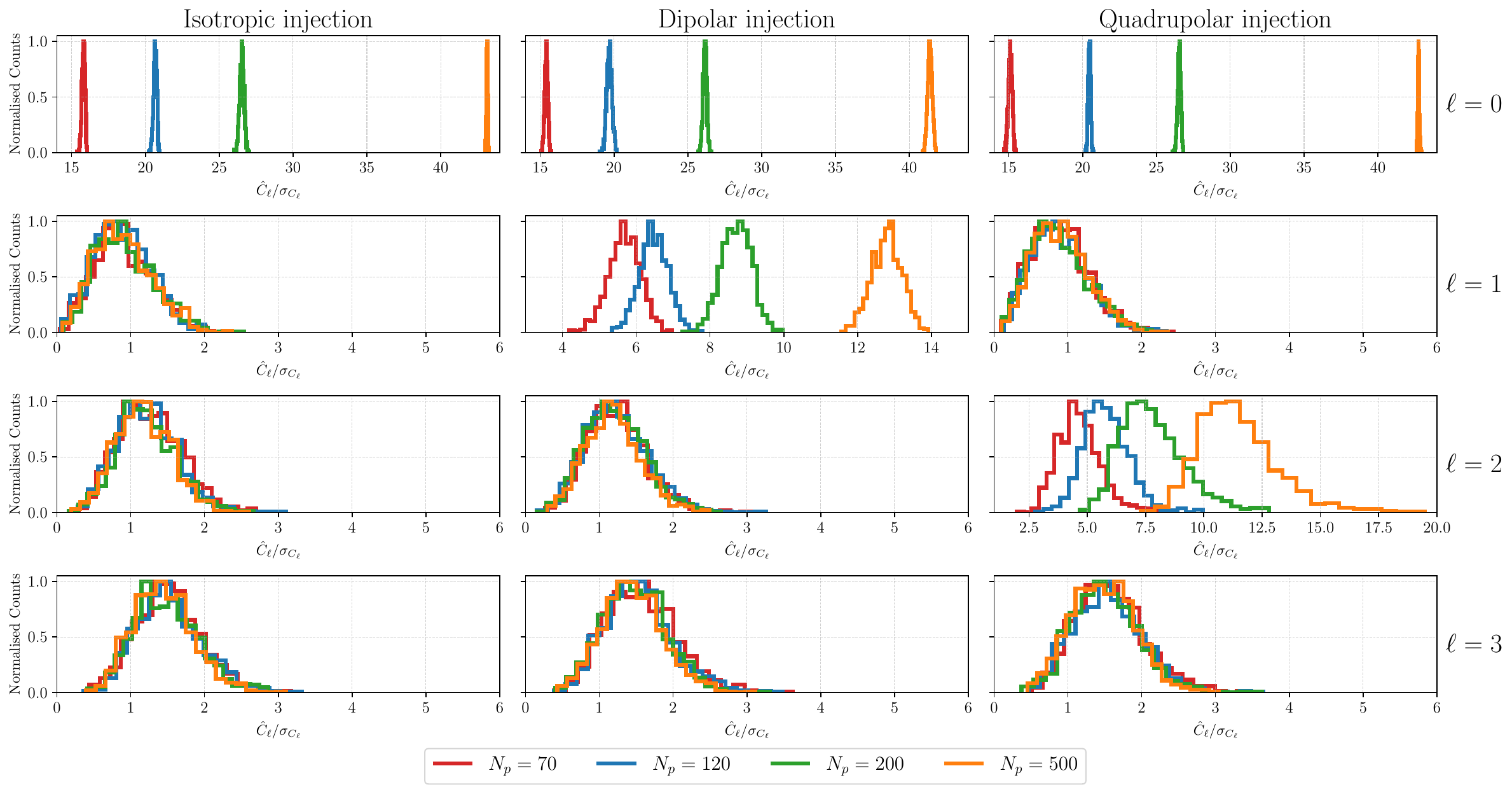}
    \caption{
    Distributions for rescaled ML $\hat C_\ell/\sigma_{C_\ell}$ recovered from $10^3$ simulated PTA datasets with varying $N_p$ and $N_{\rm side} =12$. Left to right: monopole-only, monopole+dipole, and monopole+quadrupole injected signals. 
}
    \label{fig:Cl/dCl}
\end{figure*}

\vspace{2mm}
\textbf{\textit{ Results. }} We analyse two distinct scenarios: isotropic and anisotropic GWB simulations, with various $N_p$. 
We will focus on the widely adopted angular power spectra
\begin{equation}
    C_\ell \equiv \sum_m \frac{c_{\ell m}^2}{2 \ell +1}.
\end{equation} 
Our results are $N_{\rm pix}$ (or $\Delta \hat \Omega$)-independent, as we adopt values well below the angular resolution of our PTA configurations. We use $S_h =f^{-7/3}$ as predicted by GW driven SMBHB inspirals \cite{Phinney:2001di} and $N_f = 15$.

{\it Isotropic simulations.}
For isotropic GWB realizations, we inject signals with $c_{00} = 1/\sqrt{4\pi}$ and all higher-order multipole coefficients set to zero.
Fig.~\ref{fig:Cl/dCl} (leftmost column) shows the distribution of reconstructed multipole power spectra normalized by their associated uncertainties $\sigma_{C_\ell}$, for various PTA configurations.

In the top panel, we see that with larger $N_p$, the monopole $C_0 = 1/4 \pi$ is measured with higher and higher accuracy, following the expected scaling $\sigma_{C_0} \sim 1/\sqrt{N_p}$ \cite{Babak:2024yhu} as we include autocorrelations in the analysis.
Moving to higher order multipoles, we see that the inferred $\hat  C_\ell$ are always positive, as in each realization, due to the stochastic nature of the signal, the inferred $\hat c_{\ell m}$ are non-zero. 
In other words, the distribution of recovered $\hat C_\ell$ exhibits a characteristic positive expectation value inherent to power spectrum estimators: although the underlying $c_{\ell m}$ coefficients have zero mean, they possess non-zero variance, and their quadratic combination $C_\ell$, being the sum of squares, yields strictly positive values. However, the key finding is that the ratio $\hat C_\ell/\sigma_{C_\ell}$ consistently peaks around ${\cal O}(1)$ across all higher multipoles, confirming that the induced and reconstructed anisotropies remain small compared to the measurement uncertainties, and one infers the signal to be statistically consistent with isotropy. This analysis demonstrates that cosmic variance does not induce spurious anisotropy detections. 

Our conclusion remains valid across different simulation parameters. The recovery of isotropy is independent of the sky pixelization scheme, provided the strong-signal approximation remains valid. Moreover, changing the number of pixels does not affect the results since it only redistributes the same total variance (see footnote~\ref{footnote:pixelization}). Similarly, the result remains unaffected by varying the number of pulsars $N_p$ from 70 to 500. This shows that, while having larger arrays reduces measurement uncertainties (as established in previous work~\cite{Depta:2024ykq}), a larger $N_p$ also decreases the cosmic variance because one takes more independent measurements of the sky.
All in all, we find that isotropic backgrounds are correctly identified as such, even when accounting for cosmic variance in the simulation of single data realizations.

We expect these results to extend to more realistic observational scenarios. Including instrumental noise would only increase the measurement uncertainties, making the results even more consistent with isotropy. For sparse PTA networks where pulsar-limited rather than cosmic variance dominates the error budget, the statistical inference remains equally reliable.

{\it Anisotropic simulations.}
To validate the ability of our framework to detect genuine anisotropies, we perform injections with non-zero multipole coefficients up to $\ell_{\text{max}} = 2$, i.e., a maximally dipolar signal $P(\hat{\Omega}) = (1 + \cos\theta)/4\pi$, corresponding to an injection with only non-zero $\{c_{00} = 1/(2\sqrt{\pi}), c_{10} = 1/(2\sqrt{3\pi})\}$, and a maximally quadrupolar signal $P(\hat{\Omega}) = 3\cos^2\theta/4\pi$, corresponding to $\{c_{00} = 1/(2\sqrt{\pi}), c_{20} = 1/(\sqrt{5\pi})\}$.
The corresponding results are shown in the center and right columns of Fig.~\ref{fig:Cl/dCl}.

A few comments are necessary. First, we observe that the monopole is recovered with the same significance as in the isotropic scenario. This is because different multipoles are independent, and the total (integrated) power remains unchanged under anisotropic injections, i.e., $\int {\rm d}^2 \hat \Omega P(\hat \Omega) = 1 $. Therefore, in the strong-signal regime, the uncertainty is driven solely by $N_f$ and $N_p$. 
Furthermore, for the multipoles with no injected signal, the distribution of $\hat C_\ell / \sigma_{C_\ell}$ remains of order ${\cal O}(1)$, i.e., the multipole is consistent with zero, as in the isotropic case. 
Finally, for the multipoles corresponding to the injection (dipolar: $\ell = 1$, quadrupolar: $\ell = 2$), the $C_\ell$ are measured with increasing significance as $N_p$ grows.

\vspace{2mm}
\textbf{\textit{Conclusions.}}
In this work, we have performed simulations of realistic PTA datasets by injecting GWB signals with different angular power spectra. Adopting an inference framework based on the Whittle likelihood \eqref{eq:likelihood}, we have analyzed the case of signal-dominated observations while varying the number of pulsars. Our results confirm the fundamental limitations of PTA experiments to access the underlying $C_{\ell}$ distribution. Nevertheless, the inferred parameters remain robustly consistent with the assumptions underlying each injection, and the behaviour of anisotropies due to cosmic variance follows the expected trends.
Our simulations confirm that cosmic variance cannot induce significant false anisotropic signals, provided the correct likelihood is adopted\footnote{As shown in \cite{Konstandin:2024fyo}, modeling the time-domain cross-correlations $C_{IJ}$ as Gaussian variables, rather than the square of Gaussian variables, neglects higher order correlations affecting the analysis. }. Even when individual realizations of an isotropic GWB exhibit apparent directional variations due to cosmic variance, the statistical uncertainties properly account for this effect. For an isotropic GWB, the recovered confidence intervals remain consistent with the null hypothesis of isotropy, providing a robust foundation for distinguishing astrophysical from cosmological GWB origins based on the search for anisotropies.

The cosmic variance in PTAs is analogous to the well-known CMB case after taking into account the more complicated structure of the response function arising from the three-dimensional network geometry.
In CMB temperature measurements, cosmic variance at multipole $\ell$ is fundamentally limited by the $\sim 2\ell + 1$ independent modes available on the celestial sphere. This results in large error bars, often shown as the cosmic variance `band'. For the CMB, all information we have is a 2D map of the sky, so assuming there are no limitations due to the instrument resolution, the number of measurements we have at multipole $\ell$ is simply $\sim 2 \ell + 1$. However, PTAs probe the GWB through a network of pulsar pairs, each acting as an independent detector with its unique directional sensitivity encoded in the generalized overlap reduction functions $\Gamma_{IJ,\ell m}$.
This geometric advantage allows PTAs to potentially overcome some cosmic variance limitations by increasing the number of pulsars. Each additional pulsar contributes $N_p - 1$ new correlations, providing additional independent measurements of the same underlying multipole structure. 
Therefore, we confirm that future PTA datasets could detect anisotropies as small as $C_{\ell = 1} = 9 \times 10^{-3} (1/N_f) (100/N_p) $ at $3\sigma$ C.L., where $N_f$ indicates the number of signal dominated observed frequencies, as reported in \cite{Depta:2024ykq}. 

\vspace{2mm}
\begin{acknowledgments}
We thank A.~Mitridate and J.~Romano for helpful discussions.
We also thank G. Mentasti for discussions on the maximum multipole constrained $\ell_{\rm max}$ that inspired App.~\ref{app:lmax}.
The work of MP is supported by the Comunidad de Madrid under the Programa de Atracción de Talento Investigador with number 2024-T1TEC-3134. MP acknowledges the hospitality of Imperial College London, which provided office space during some parts of this project.
\end{acknowledgments}

\appendix

\onecolumngrid

\section{Maximum Number of Multipoles Constrained with $N_p$ Pulsars}
\label{app:lmax}

We estimate the number of independent multipole moments $c_{\ell m}$ that can be constrained from GWB anisotropies using $N_p$ pulsars. The estimator for the multipole coefficients can be written as
\begin{equation}
	c(f) \sum_{\ell =0}^{\ell_{\rm max}}
\sum_{m =-\ell}^{\ell} c_{\ell m}  
\Gamma_{IJ, \ell m} = \langle \tilde d_I \tilde d_J \rangle,
\end{equation}
where we have factored out a frequency-dependent amplitude $c(f)$. 
A formal solution of this linear system involving $(\ell_{\rm max}+1)^2$ unknowns can be easily obtained. 
We rewrite this system of equations indicating as $\alpha$ and $\beta$ the index of pulsar pairs and $c_{\ell m}$ coefficients, to get
\begin{equation}
    c_{\alpha}
\propto 
\Gamma_{\alpha \beta}^{-1} D_{\alpha}.
\end{equation}
The matrix inversion requires a square matrix of full rank, indicating the maximal number of ${c_{\ell m}}'s$ which can be inferred is given by the length of $\{ \alpha \}$, i.e.\ the number of pulsar pairs $N_{\rm pairs}$.
Given $N_p$ pulsars, the number of independent cross-correlation pairs is
$N_{\rm pairs} = {N_p (N_p - 1)}/{2}$.
Assuming that only cross-correlations contribute significantly (neglecting autocorrelations relevant for low $\ell$, see \cite{Depta:2024ykq}), and considering a single frequency bin for simplicity, the estimation of $c_{\ell m}$ can be schematically written as
\begin{equation}
\begin{pmatrix}
c_{1\,0} \\
c_{1\,-1} \\
\vdots \\
c_{\lmax,\lmax}
\end{pmatrix}
=
\begin{pmatrix}
 & & \\
 & \Gamma_{\alpha \beta}^{-1}   & \\
 & & 
\end{pmatrix}
\begin{pmatrix}
D_{1\,2} \\
D_{1\,3} \\
\vdots \\
D_{N_p\,N_p-1}
\end{pmatrix}.	
\end{equation}
We have verified explicitly that the rank of $ \Gamma_{\alpha \beta}^{-1}$ is generically equal to the number of pulsar pairs,
indicating that the number of multipole coefficients that can be independently constrained scales as
\begin{equation}
	(\lmax+1)^2 \sim \frac{N_p (N_p-1)}{2}
	\quad \Rightarrow \quad 
	\lmax \sim N_p.
\end{equation}  
This scaling differs from the $\lmax \sim \sqrt{N_p}$ estimate quoted in \cite{NANOGrav:2023tcn,Romano:2016dpx}.
We provide a physical interpretation for this scaling argument in the main text.

\section{Model validation through probability - probability (P-P) plots}\label{app:pp-plot}

\begin{figure*}[h]
    \centering
\includegraphics[width=.32\linewidth]{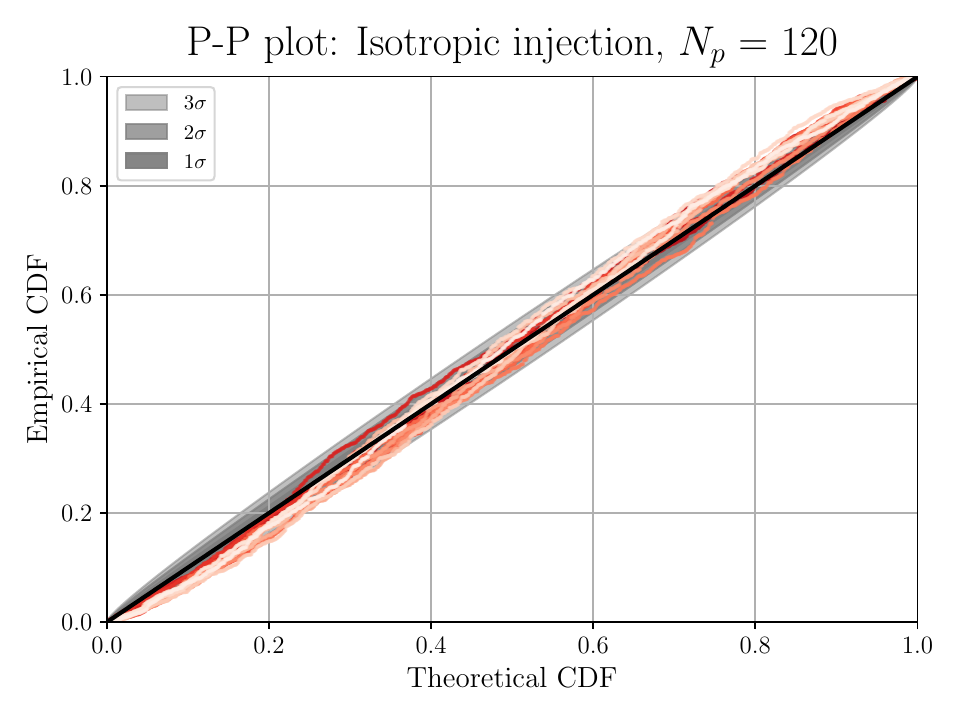}
\includegraphics[width=.32\linewidth]{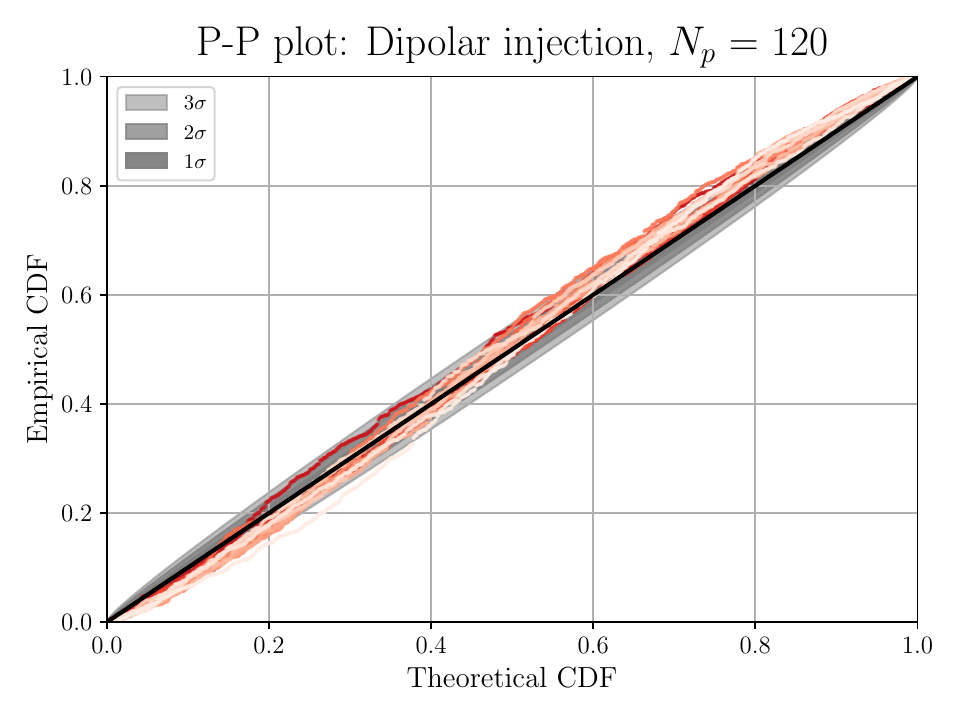}
\includegraphics[width=.32\linewidth]{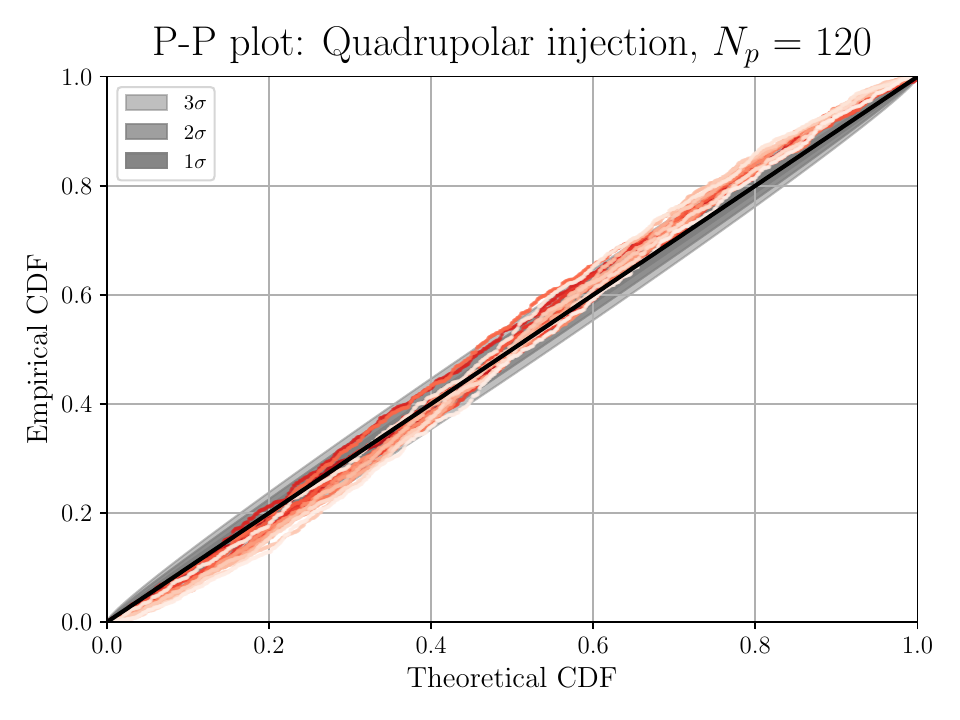}
\includegraphics[width=.32\linewidth]{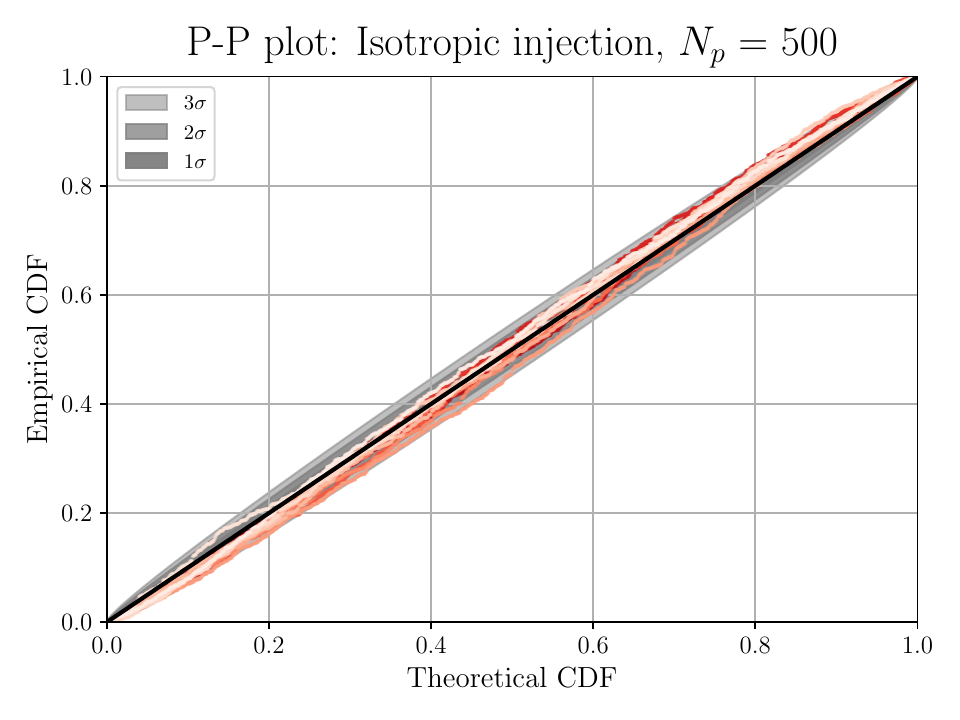}
\includegraphics[width=.32\linewidth]{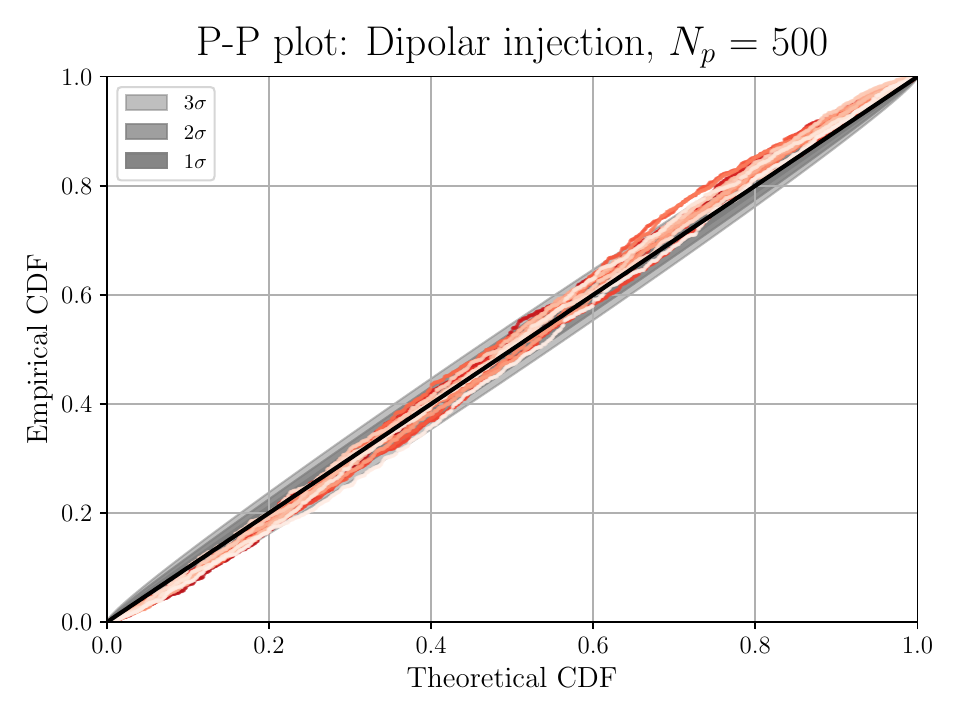}
\includegraphics[width=.32\linewidth]{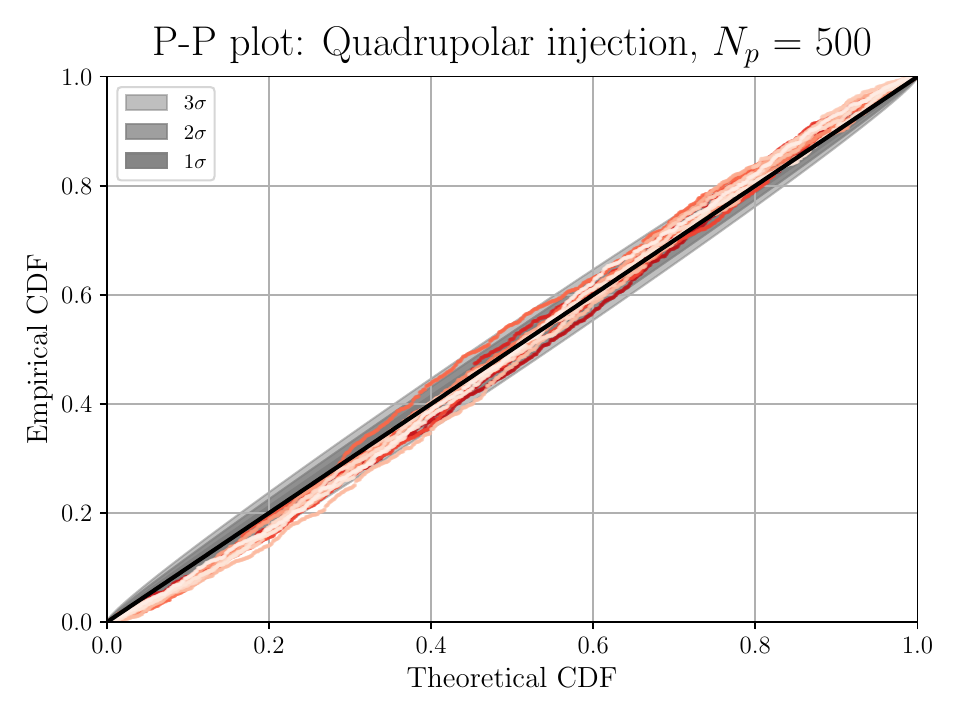}
\caption{
P-P plots for all $c_{\ell m}$ up to $\ell_{\text{max}} = 3$ (different red tones), using $10^3$ realizations with $N_p = 120$ (top row) and $N_p = 500$ (bottom row). From left to right: monopole-only, monopole+dipole, and monopole+quadrupole injections.
The shaded grey region denotes the $1, 2, 3 \, \sigma$ confidence band expected from statistical fluctuations, see the main text for its definition. The close match to the diagonal confirms the validity of the inference and FIM-based uncertainties.
}
\label{fig:PP plot}
\end{figure*}

We validate our inference procedure, as well as the accuracy of the uncertainty estimates obtained using the FIM, by presenting the probability-probability (P-P) plot shown in Fig.~\ref{fig:PP plot}.

The P--P plot compares the cumulative distribution function (CDF) of the inferred quantile for the injected $c_{\ell m}$ values in the posterior, with the uniform distribution expected under correct calibration, providing a diagnostic of the statistical consistency of the inferred posteriors. 
To construct it, we generate $1000$ independent data realizations. For each realization, we perform the full inference and determine the quantile at which the injected $c_{\ell m}$ value lies within the corresponding posterior distribution. The resulting set of quantiles is then sorted in ascending order and plotted against the theoretical expectation for ordered quantiles, which follows an equally spaced sequence between $0$ and $1$. 
The $1\sigma$ sampling uncertainty on the P--P plot is computed assuming binomial statistics: for a given theoretical cumulative probability $p$, the standard deviation is 
\begin{equation}
    \sigma(p) = \sqrt{\frac{p(1-p)}{N}},
\end{equation}
where $N$ is the total number of realizations. This uncertainty band,  shown as a gray shaded region around the diagonal, quantifies the expected statistical scatter for a finite sample of realizations.
See App. A of~\cite{Franciolini:2025leq}
for more explanations and a dictionary on how to interpret P-P plots.

In Fig.~\ref{fig:PP plot}, from left to right, we display the P-P plot for each $c_{\ell m}$ coefficient up to $\ell_{\text{max}} = 3$, corresponding respectively to injections with: monopole only, monopole plus maximal dipole, and monopole plus maximal quadrupole. All results are based on $10^3$ realizations, each using $N_p = 120$ and $N_p = 500$ pulsars. We find excellent agreement with the theoretical expectation in all cases, confirming both the validity of the inference method and the reliability of the associated uncertainty estimates.

\begin{figure*}[b]
\centering
\includegraphics[width=1\linewidth]{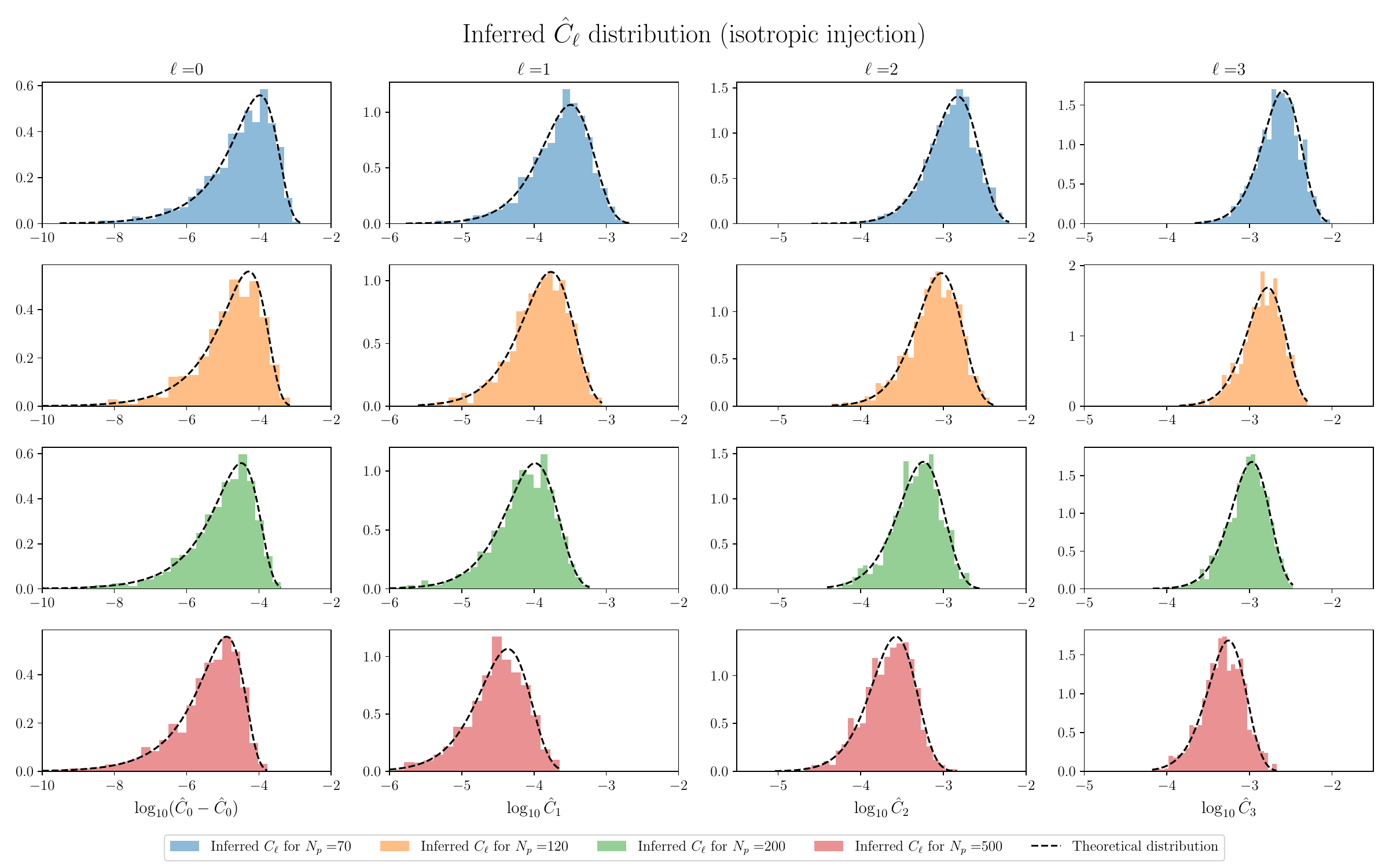}
\caption{
Distribution of the maximum likelihood $\hat C_\ell$ computed with $1000$ realizations of the injected signal. 
}
\label{fig:hatCl}
\end{figure*}

\section{Distribution of $\hat C_\ell$}\label{app:C_l_dist}

In Fig.~\ref{fig:hatCl} we show the distribution of the ML estimators $\hat C_\ell$ obtained performing 1000 realizations. We assume the same parameters as in Fig.~\ref{fig:Cl/dCl}. Each color (row) corresponds to a different number of pulsars, and each column corresponds to a different multiple index $\ell$. The simulations assume an isotropic GWB. We remove the mean from the $\ell = 0$ result.
Superimposed to the histograms, we show in black dashed line the expected distribution of $C_\ell$, built assuming a Gaussian $d_{I,i}$ for an isotropic signal. This reads \cite{Franciolini:2025leq}
\begin{equation}
p(D|P) = \frac{N_s^{N_s} D^{N_s-1}}{\Gamma(N_s)}\frac{e^{-N_s D /P}}{P^{N_s}}\,,
\label{e:p(Phat|P)}
\end{equation}
where $D = \sum_\ell |\hat c_{\ell m}|^2/ (2 \ell+1)$, the variance is 
$P = \text{mean}(\hat C_\ell)$ across all the realizations, and $N_s = (2 \ell + 1)/2$ is the number of effective degrees of freedom in the sum at each $\ell$. 
This distribution exactly matches a $\chi^2$ distribution with the appropriate degrees of freedom $k = 2\ell + 1$, as expected for a variable built as the sum of $2 \ell + 1$ Gaussian variables.

\twocolumngrid
\bibliographystyle{utphys}
\bibliography{main}

\providecommand{\href}[2]{#2}\begingroup\raggedright\begin{thebibliography}{10}

\bibitem{Hellings:1983fr}
R.~w. Hellings and G.~s. Downs, ``{UPPER LIMITS ON THE ISOTROPIC GRAVITATIONAL
  RADIATION BACKGROUND FROM PULSAR TIMING ANALYSIS},''
  \href{http://dx.doi.org/10.1086/183954}{{\em Astrophys. J. Lett.} {\bf 265}
  (1983)  L39--L42}.

\bibitem{NANOGrav:2023gor}
{\bf NANOGrav} Collaboration, G.~Agazie {\em et al.}, ``{The NANOGrav 15 yr
  Data Set: Evidence for a Gravitational-wave Background},''
  \href{http://dx.doi.org/10.3847/2041-8213/acdac6}{{\em Astrophys. J. Lett.}
  {\bf 951} (2023) no.~1, L8}, \href{http://arxiv.org/abs/2306.16213}{{\tt
  arXiv:2306.16213 [astro-ph.HE]}}.

\bibitem{Antoniadis:2023ott}
{\bf EPTA, InPTA:} Collaboration, J.~Antoniadis {\em et al.}, ``{The second
  data release from the European Pulsar Timing Array - III. Search for
  gravitational wave signals},''
  \href{http://dx.doi.org/10.1051/0004-6361/202346844}{{\em Astron. Astrophys.}
  {\bf 678} (2023)  A50}, \href{http://arxiv.org/abs/2306.16214}{{\tt
  arXiv:2306.16214 [astro-ph.HE]}}.

\bibitem{Reardon:2023gzh}
D.~J. Reardon {\em et al.}, ``{Search for an Isotropic Gravitational-wave
  Background with the Parkes Pulsar Timing Array},''
  \href{http://dx.doi.org/10.3847/2041-8213/acdd02}{{\em Astrophys. J. Lett.}
  {\bf 951} (2023) no.~1, L6}, \href{http://arxiv.org/abs/2306.16215}{{\tt
  arXiv:2306.16215 [astro-ph.HE]}}.

\bibitem{Xu:2023wog}
H.~Xu {\em et al.}, ``{Searching for the Nano-Hertz Stochastic Gravitational
  Wave Background with the Chinese Pulsar Timing Array Data Release I},''
  \href{http://dx.doi.org/10.1088/1674-4527/acdfa5}{{\em Res. Astron.
  Astrophys.} {\bf 23} (2023) no.~7, 075024},
  \href{http://arxiv.org/abs/2306.16216}{{\tt arXiv:2306.16216 [astro-ph.HE]}}.

\bibitem{Madge:2023cak}
E.~Madge, E.~Morgante, C.~Puchades-Ib\'a\~nez, N.~Ramberg, W.~Ratzinger,
  S.~Schenk, and P.~Schwaller, ``{Primordial gravitational waves in the
  nano-Hertz regime and PTA data \textemdash{} towards solving the GW inverse
  problem},'' \href{http://dx.doi.org/10.1007/JHEP10(2023)171}{{\em JHEP} {\bf
  10} (2023)  171}, \href{http://arxiv.org/abs/2306.14856}{{\tt
  arXiv:2306.14856 [hep-ph]}}.

\bibitem{NANOGrav:2023hvm}
{\bf NANOGrav} Collaboration, A.~Afzal {\em et al.}, ``{The NANOGrav 15 yr Data
  Set: Search for Signals from New Physics},''
  \href{http://dx.doi.org/10.3847/2041-8213/acdc91}{{\em Astrophys. J. Lett.}
  {\bf 951} (2023) no.~1, L11}, \href{http://arxiv.org/abs/2306.16219}{{\tt
  arXiv:2306.16219 [astro-ph.HE]}}.

\bibitem{Antoniadis:2023xlr}
{\bf EPTA, InPTA} Collaboration, J.~Antoniadis {\em et al.}, ``{The second data
  release from the European Pulsar Timing Array - IV. Implications for massive
  black holes, dark matter, and the early Universe},''
  \href{http://dx.doi.org/10.1051/0004-6361/202347433}{{\em Astron. Astrophys.}
  {\bf 685} (2024)  A94}, \href{http://arxiv.org/abs/2306.16227}{{\tt
  arXiv:2306.16227 [astro-ph.CO]}}.

\bibitem{Figueroa:2023zhu}
D.~G. Figueroa, M.~Pieroni, A.~Ricciardone, and P.~Simakachorn, ``{Cosmological
  Background Interpretation of Pulsar Timing Array Data},''
  \href{http://arxiv.org/abs/2307.02399}{{\tt arXiv:2307.02399 [astro-ph.CO]}}.

\bibitem{Ellis:2023oxs}
J.~Ellis, M.~Fairbairn, G.~Franciolini, G.~H\"utsi, A.~Iovino, M.~Lewicki,
  M.~Raidal, J.~Urrutia, V.~Vaskonen, and H.~Veerm\"ae, ``{What is the source
  of the PTA GW signal?},''
  \href{http://dx.doi.org/10.1103/PhysRevD.109.023522}{{\em Phys. Rev. D} {\bf
  109} (2024) no.~2, 023522}, \href{http://arxiv.org/abs/2308.08546}{{\tt
  arXiv:2308.08546 [astro-ph.CO]}}.

\bibitem{Mingarelli:2013dsa}
C.~M.~F. Mingarelli, T.~Sidery, I.~Mandel, and A.~Vecchio, ``{Characterizing
  gravitational wave stochastic background anisotropy with pulsar timing
  arrays},'' \href{http://dx.doi.org/10.1103/PhysRevD.88.062005}{{\em Phys.
  Rev. D} {\bf 88} (2013) no.~6, 062005},
  \href{http://arxiv.org/abs/1306.5394}{{\tt arXiv:1306.5394 [astro-ph.HE]}}.

\bibitem{Taylor:2013esa}
S.~R. Taylor and J.~R. Gair, ``{Searching For Anisotropic Gravitational-wave
  Backgrounds Using Pulsar Timing Arrays},''
  \href{http://dx.doi.org/10.1103/PhysRevD.88.084001}{{\em Phys. Rev. D} {\bf
  88} (2013)  084001}, \href{http://arxiv.org/abs/1306.5395}{{\tt
  arXiv:1306.5395 [gr-qc]}}.

\bibitem{Mingarelli:2017fbe}
C.~M.~F. Mingarelli, T.~J.~W. Lazio, A.~Sesana, J.~E. Greene, J.~A. Ellis,
  C.-P. Ma, S.~Croft, S.~Burke-Spolaor, and S.~R. Taylor, ``{The Local
  Nanohertz Gravitational-Wave Landscape From Supermassive Black Hole
  Binaries},'' \href{http://dx.doi.org/10.1038/s41550-017-0299-6}{{\em Nature
  Astron.} {\bf 1} (2017) no.~12, 886--892},
  \href{http://arxiv.org/abs/1708.03491}{{\tt arXiv:1708.03491 [astro-ph.GA]}}.

\bibitem{NANOGrav:2023tcn}
{\bf NANOGrav} Collaboration, G.~Agazie {\em et al.}, ``{The NANOGrav 15 yr
  Data Set: Search for Anisotropy in the Gravitational-wave Background},''
  \href{http://dx.doi.org/10.3847/2041-8213/acf4fd}{{\em Astrophys. J. Lett.}
  {\bf 956} (2023) no.~1, L3}, \href{http://arxiv.org/abs/2306.16221}{{\tt
  arXiv:2306.16221 [astro-ph.HE]}}.

\bibitem{Sah:2024oyg}
M.~R. Sah, S.~Mukherjee, V.~Saeedzadeh, A.~Babul, M.~Tremmel, and T.~R. Quinn,
  ``{Imprints of Supermassive Black Hole Evolution on the Spectral and Spatial
  Anisotropy of Nano-Hertz Stochastic Gravitational-Wave Background},''
  \href{http://arxiv.org/abs/2404.14508}{{\tt arXiv:2404.14508 [astro-ph.CO]}}.

\bibitem{Sah:2024etc}
M.~R. Sah and S.~Mukherjee, ``{Discovering the Cosmic Evolution of Supermassive
  Black Holes using Nano-Hertz Gravitational Waves and Galaxy Surveys},''
  \href{http://arxiv.org/abs/2407.11669}{{\tt arXiv:2407.11669 [astro-ph.CO]}}.

\bibitem{Nay:2023pwu}
J.~Nay, K.~K. Boddy, T.~L. Smith, and C.~M.~F. Mingarelli, ``{Harmonic analysis
  for pulsar timing arrays},''
  \href{http://dx.doi.org/10.1103/PhysRevD.110.044062}{{\em Phys. Rev. D} {\bf
  110} (2024) no.~4, 044062}, \href{http://arxiv.org/abs/2306.06168}{{\tt
  arXiv:2306.06168 [gr-qc]}}.

\bibitem{Pitrou:2024scp}
C.~Pitrou and G.~Cusin, ``{Mitigating cosmic variance in the Hellings-Downs
  curve: a Cosmic Microwave Background analogy},''
  \href{http://arxiv.org/abs/2412.12073}{{\tt arXiv:2412.12073 [gr-qc]}}.

\bibitem{Allen:2022dzg}
B.~Allen, ``{Variance of the Hellings-Downs correlation},''
  \href{http://dx.doi.org/10.1103/PhysRevD.107.043018}{{\em Phys. Rev. D} {\bf
  107} (2023) no.~4, 043018}, \href{http://arxiv.org/abs/2205.05637}{{\tt
  arXiv:2205.05637 [gr-qc]}}.

\bibitem{Grimm:2024lfj}
N.~Grimm, M.~Pijnenburg, G.~Cusin, and C.~Bonvin, ``{The impact of large-scale
  galaxy clustering on the variance of the Hellings-Downs correlation},''
  \href{http://arxiv.org/abs/2404.05670}{{\tt arXiv:2404.05670 [astro-ph.CO]}}.

\bibitem{Cusin:2025xle}
G.~Cusin, C.~Pitrou, M.~Pijnenburg, and A.~Sesana, ``{Measuring anisotropies in
  the PTA band with cross-correlations},''
  \href{http://arxiv.org/abs/2502.17401}{{\tt arXiv:2502.17401 [gr-qc]}}.

\bibitem{Agarwal:2024hlj}
D.~Agarwal and J.~D. Romano, ``{Cosmic variance of the Hellings and Downs
  correlation for ensembles of universes having non-zero angular power
  spectra},'' \href{http://arxiv.org/abs/2404.08574}{{\tt arXiv:2404.08574
  [gr-qc]}}.

\bibitem{Becsy:2022pnr}
B.~B\'ecsy, N.~J. Cornish, and L.~Z. Kelley, ``{Exploring Realistic Nanohertz
  Gravitational-wave Backgrounds},''
  \href{http://dx.doi.org/10.3847/1538-4357/aca1b2}{{\em Astrophys. J.} {\bf
  941} (2022) no.~2, 119}, \href{http://arxiv.org/abs/2207.01607}{{\tt
  arXiv:2207.01607 [astro-ph.HE]}}.

\bibitem{Sato-Polito:2023spo}
G.~Sato-Polito and M.~Kamionkowski, ``{Exploring the spectrum of stochastic
  gravitational-wave anisotropies with pulsar timing arrays},''
  \href{http://dx.doi.org/10.1103/PhysRevD.109.123544}{{\em Phys. Rev. D} {\bf
  109} (2024) no.~12, 123544}, \href{http://arxiv.org/abs/2305.05690}{{\tt
  arXiv:2305.05690 [astro-ph.CO]}}.

\bibitem{Lemke:2024cdu}
A.-M. Lemke, A.~Mitridate, and K.~A. Gersbach, ``{Detecting Gravitational Wave
  Anisotropies from Supermassive Black Hole Binaries},''
  \href{http://arxiv.org/abs/2407.08705}{{\tt arXiv:2407.08705 [astro-ph.HE]}}.

\bibitem{Gersbach:2024hcc}
K.~A. Gersbach, S.~R. Taylor, P.~M. Meyers, and J.~D. Romano, ``{Spatial and
  Spectral Characterization of the Gravitational-wave Background with the PTA
  Optimal Statistic},'' \href{http://arxiv.org/abs/2406.11954}{{\tt
  arXiv:2406.11954 [astro-ph.IM]}}.

\bibitem{Raidal:2024tui}
J.~Raidal, J.~Urrutia, V.~Vaskonen, and H.~Veerm{\"a}e, ``{Statistics of the
  supermassive black hole gravitational wave background anisotropy},''
  \href{http://arxiv.org/abs/2411.19692}{{\tt arXiv:2411.19692 [astro-ph.CO]}}.

\bibitem{Moreschi:2025qtm}
B.~E. Moreschi, S.~Valtolina, A.~Sesana, G.~Shaifullah, M.~Falxa, L.~Speri,
  D.~Izquierdo-Villalba, and A.~Chalumeau, ``{Dissecting the nanoHz
  gravitational wave sky: frequency-correlated anisotropy induced by eccentric
  supermassive black hole binaries},''
  \href{http://arxiv.org/abs/2506.14882}{{\tt arXiv:2506.14882 [astro-ph.GA]}}.

\bibitem{Allen:1996gp}
B.~Allen and A.~C. Ottewill, ``{Detection of anisotropies in the gravitational
  wave stochastic background},''
  \href{http://dx.doi.org/10.1103/PhysRevD.56.545}{{\em Phys. Rev. D} {\bf 56}
  (1997)  545--563}, \href{http://arxiv.org/abs/gr-qc/9607068}{{\tt
  arXiv:gr-qc/9607068}}.

\bibitem{Anholm:2008wy}
M.~Anholm, S.~Ballmer, J.~D.~E. Creighton, L.~R. Price, and X.~Siemens,
  ``{Optimal strategies for gravitational wave stochastic background searches
  in pulsar timing data},''
  \href{http://dx.doi.org/10.1103/PhysRevD.79.084030}{{\em Phys. Rev. D} {\bf
  79} (2009)  084030}, \href{http://arxiv.org/abs/0809.0701}{{\tt
  arXiv:0809.0701 [gr-qc]}}.

\bibitem{Taylor:2015udp}
S.~R. Taylor {\em et al.}, ``{Limits on anisotropy in the nanohertz stochastic
  gravitational-wave background},''
  \href{http://dx.doi.org/10.1103/PhysRevLett.115.041101}{{\em Phys. Rev.
  Lett.} {\bf 115} (2015) no.~4, 041101},
  \href{http://arxiv.org/abs/1506.08817}{{\tt arXiv:1506.08817 [astro-ph.HE]}}.

\bibitem{Taylor:2020zpk}
S.~R. Taylor, R.~van Haasteren, and A.~Sesana, ``{From Bright Binaries To Bumpy
  Backgrounds: Mapping Realistic Gravitational Wave Skies With Pulsar-Timing
  Arrays},'' \href{http://dx.doi.org/10.1103/PhysRevD.102.084039}{{\em Phys.
  Rev. D} {\bf 102} (2020) no.~8, 084039},
  \href{http://arxiv.org/abs/2006.04810}{{\tt arXiv:2006.04810 [astro-ph.IM]}}.

\bibitem{Ali-Haimoud:2020ozu}
Y.~Ali-Ha\"\i{}moud, T.~L. Smith, and C.~M.~F. Mingarelli, ``{Fisher formalism
  for anisotropic gravitational-wave background searches with pulsar timing
  arrays},'' \href{http://dx.doi.org/10.1103/PhysRevD.102.122005}{{\em Phys.
  Rev. D} {\bf 102} (2020) no.~12, 122005},
  \href{http://arxiv.org/abs/2006.14570}{{\tt arXiv:2006.14570 [gr-qc]}}.

\bibitem{Ali-Haimoud:2020iyz}
Y.~Ali-Ha\"\i{}moud, T.~L. Smith, and C.~M.~F. Mingarelli, ``{Insights into
  searches for anisotropies in the nanohertz gravitational-wave background},''
  \href{http://dx.doi.org/10.1103/PhysRevD.103.042009}{{\em Phys. Rev. D} {\bf
  103} (2021) no.~4, 042009}, \href{http://arxiv.org/abs/2010.13958}{{\tt
  arXiv:2010.13958 [gr-qc]}}.

\bibitem{Pol:2022sjn}
N.~Pol, S.~R. Taylor, and J.~D. Romano, ``{Forecasting Pulsar Timing Array
  Sensitivity to Anisotropy in the Stochastic Gravitational Wave Background},''
  \href{http://dx.doi.org/10.3847/1538-4357/ac9836}{{\em Astrophys. J.} {\bf
  940} (2022) no.~2, 173}, \href{http://arxiv.org/abs/2206.09936}{{\tt
  arXiv:2206.09936 [astro-ph.HE]}}.

\bibitem{Depta:2024ykq}
P.~F. Depta, V.~Domcke, G.~Franciolini, and M.~Pieroni, ``{Pulsar timing array
  sensitivity to anisotropies in the gravitational wave background},''
  \href{http://dx.doi.org/10.1103/PhysRevD.111.083039}{{\em Phys. Rev. D} {\bf
  111} (2025) no.~8, 083039}, \href{http://arxiv.org/abs/2407.14460}{{\tt
  arXiv:2407.14460 [astro-ph.CO]}}.

\bibitem{Romano:2016dpx}
J.~D. Romano and N.~J. Cornish, ``{Detection methods for stochastic
  gravitational-wave backgrounds: a unified treatment},''
  \href{http://dx.doi.org/10.1007/s41114-017-0004-1}{{\em Living Rev. Rel.}
  {\bf 20} (2017) no.~1, 2}, \href{http://arxiv.org/abs/1608.06889}{{\tt
  arXiv:1608.06889 [gr-qc]}}.

\bibitem{Blandford:1984xwb}
R.~Blandford, R.~Narayan, and R.~W. Romani, ``{Arrival-time analysis for a
  millisecond pulsar},'' \href{http://dx.doi.org/10.1007/BF02714466}{{\em J.
  Astrophys. Astron.} {\bf 5} (1984) no.~4, 369--388}.

\bibitem{Hazboun:2019vhv}
J.~S. Hazboun, J.~D. Romano, and T.~L. Smith, ``{Realistic sensitivity curves
  for pulsar timing arrays},''
  \href{http://dx.doi.org/10.1103/PhysRevD.100.104028}{{\em Phys. Rev. D} {\bf
  100} (2019) no.~10, 104028}, \href{http://arxiv.org/abs/1907.04341}{{\tt
  arXiv:1907.04341 [gr-qc]}}.

\bibitem{Babak:2024yhu}
S.~Babak, M.~Falxa, G.~Franciolini, and M.~Pieroni, ``{Forecasting the
  sensitivity of Pulsar Timing Arrays to gravitational wave backgrounds},''
  \href{http://arxiv.org/abs/2404.02864}{{\tt arXiv:2404.02864 [astro-ph.CO]}}.

\bibitem{Kehagias:2024plp}
A.~Kehagias and A.~Riotto, ``{The PTA Hellings and Downs correlation unmasked
  by symmetries},'' \href{http://dx.doi.org/10.1088/1475-7516/2024/06/059}{{\em
  JCAP} {\bf 06} (2024)  059}, \href{http://arxiv.org/abs/2401.10680}{{\tt
  arXiv:2401.10680 [gr-qc]}}.

\bibitem{Gorski:2004by}
K.~M. G{\'o}rski, E.~Hivon, A.~J. Banday, B.~D. Wandelt, F.~K. Hansen,
  M.~Reinecke, and M.~Bartelman, ``{HEALPix - A Framework for high resolution
  discretization, and fast analysis of data distributed on the sphere},''
  \href{http://dx.doi.org/10.1086/427976}{{\em Astrophys. J.} {\bf 622} (2005)
  759--771}, \href{http://arxiv.org/abs/astro-ph/0409513}{{\tt
  arXiv:astro-ph/0409513}}.

\bibitem{Allen:2024bnk}
B.~Allen, ``{Pulsar Timing Array Harmonic Analysis and Source Angular
  Correlations},'' \href{http://arxiv.org/abs/2404.05677}{{\tt arXiv:2404.05677
  [gr-qc]}}.

\bibitem{Moran1951HypothesisTI}
P.~artist Moran and P.~Whittle, ``Hypothesis testing in time series
  analysis.,''
\newblock 1951.
\newblock \url{https://api.semanticscholar.org/CorpusID:125739077}.

\bibitem{Contaldi:2020rht}
C.~R. Contaldi, M.~Pieroni, A.~I. Renzini, G.~Cusin, N.~Karnesis, M.~Peloso,
  A.~Ricciardone, and G.~Tasinato, ``{Maximum likelihood map-making with the
  Laser Interferometer Space Antenna},''
  \href{http://dx.doi.org/10.1103/PhysRevD.102.043502}{{\em Phys. Rev. D} {\bf
  102} (2020) no.~4, 043502}, \href{http://arxiv.org/abs/2006.03313}{{\tt
  arXiv:2006.03313 [astro-ph.CO]}}.

\bibitem{Bond:1998zw}
J.~R. Bond, A.~H. Jaffe, and L.~Knox, ``{Estimating the power spectrum of the
  cosmic microwave background},''
  \href{http://dx.doi.org/10.1103/PhysRevD.57.2117}{{\em Phys. Rev. D} {\bf 57}
  (1998)  2117--2137}, \href{http://arxiv.org/abs/astro-ph/9708203}{{\tt
  arXiv:astro-ph/9708203}}.

\bibitem{Franciolini:2025leq}
G.~Franciolini, M.~Pieroni, A.~Ricciardone, and J.~D. Romano, ``{Likelihoods
  for Stochastic Gravitational Wave Background Data Analysis},''
  \href{http://arxiv.org/abs/2505.24695}{{\tt arXiv:2505.24695 [gr-qc]}}.

\bibitem{Coe:2009xf}
D.~Coe, ``{Fisher Matrices and Confidence Ellipses: A Quick-Start Guide and
  Software},'' \href{http://arxiv.org/abs/0906.4123}{{\tt arXiv:0906.4123
  [astro-ph.IM]}}.

\bibitem{Phinney:2001di}
E.~S. Phinney, ``{A Practical theorem on gravitational wave backgrounds},''
  \href{http://arxiv.org/abs/astro-ph/0108028}{{\tt arXiv:astro-ph/0108028}}.

\bibitem{Konstandin:2024fyo}
T.~Konstandin, A.-M. Lemke, A.~Mitridate, and E.~Perboni, ``{The impact of
  cosmic variance on PTAs anisotropy searches},''
  \href{http://dx.doi.org/10.1088/1475-7516/2025/04/059}{{\em JCAP} {\bf 04}
  (2025)  059}, \href{http://arxiv.org/abs/2408.07741}{{\tt arXiv:2408.07741
  [astro-ph.CO]}}.

\end{thebibliography}\endgroup

\end{document}